\documentclass{PoS}
\usepackage[utf8]{inputenc}
\usepackage{amssymb, amsmath}
\usepackage[utf8]{inputenc}
\usepackage[T1]{fontenc}

\title{Production of t$\gamma$, tZ and tH via Flavour Changing Neutral Currents}

\ShortTitle{Production of t$\gamma$, tZ and tH via Flavour Changing Neutral Currents.}

\author{\speaker{Artur Amorim} $^{a,b}$, Nuno Filipe Castro $^{a,b}$, Juan Pedro Araque $^{b}$, José Santiago Perez $^{c}$, Rui Santos $^{d}$\\
        \footnote{Artur Amorim is funded by FCT and Fundação Calouste Gulbenkian through the grants FEDER/COMPETE-QREN, FCT, Portugal (ref. IF/00050/2013/CP1172/CT0002) and Programa de Estímulo à Investigação.} \\
        $^{a}$ Departamento de Física e Astronomia, Faculdade de Ciências. Universidade do Porto,\\
        4169-007 Porto, Portugal \\
        $^{b}$ LIP, Departamento de Física, Escola de Ciências, Universidade do Minho,\\
        4710-057 Braga, Portugal \\
        $^{c}$ CAFPE and Departamento de Física Teórica y del Cosmos, Universidad de Granada, \\
        E-18071 Granada, Spain \\
        $^{d}$ Instituto Superior de Engenharia de Lisboa (ISEL), \\
        1959-007 Lisboa, Portugal \\
        
        E-mail: \email{artur@lip.pt, nfcastro@lip.pt, jaraque@lip.pt, jsantiago@ugr.es, rasantos@fc.ul.pt}}

\abstract{A UFO model describing general top quark Flavour Changing Neutral Currents is presented. We use it to study t$\gamma$, tH and tZ production via FCNCs anomalous couplings at the Large Hadron Collider, in particular how the distributions of physical observables depend on the anomalous couplings. A sensitivity study of the Large Hadron Collider experiments to tZ production via FCNC in its second stage of operation is also performed.}

\FullConference{8th International Workshop on Top Quark Physics, TOP2015\\
		14-18 September, 2015\\
		Ischia, Italy}

\begin{document}

\section{Introduction}
In the Standard Model (SM) top quark Flavour Changing Neutral Current (FCNC) processes are forbidden at tree level and are highly suppressed at higher orders due to the GIM mechanism. In Beyond Standard Model physics the top FCNC branching ratios can be enhanced by several orders of magnitude, providing an optimal scenario to search new physics.

In order to experimentally search for these processes it is useful to have a parameterisation and simulation of the most general top quark FCNC Lagrangian. Here we present a UFO model \cite{ufo} that attempts to perform these tasks as well as some results obtained with it.

\section{The UFO Model Lagrangian}

We have created a UFO model based on the most general effective Lagrangian including terms of mass dimension six. Dimension-five gauge invariant operators are not considered because they violate total lepton and baryon number conservation. The most general dimension-six gauge invariant lagrangian can be written as \cite{buchmuller} 

\begin{align}
    \mathcal{L}^{(6)} = \frac{1}{\Lambda^{2}} \sum_{x} C_{x}\mathcal{O}_{x} + h.c. \quad \text{,}
\end{align}
with $C_{x}$ a complex number, $\mathcal{O}_{x}$ a dimension-six gauge invariant operator and $\Lambda$ is the New Physics scale. In \cite{juan-top-anomalous,juan-top-higgs-anomalous} a list containing a complete set of non-redundant operators $\mathcal{O}_{x}$ is presented. Using that list the following lagrangian is obtained
\begin{align}
        &\mathcal{L}_{EFT}=\sum_{q=u,c} [ \frac{g_{s}}{2m_{t}} \bar{q} T^{a} \sigma^{\mu \nu} (\zeta^{L}_{qt}P_{L} + \zeta^{R}_{qt}P_{R}) t G^{a}_{\mu \nu } + \frac{e}{2m_{t}} \bar{q} \sigma^{\mu \nu}(\lambda^{L}_{qt}P_{L} + \lambda^{R}_{qt}P_{R}) t A_{\mu \nu } + \notag  \\ + &\frac{g_{w}}{2c_{w}} \bar{q} \gamma^{\mu} (X^{L}_{qt}P_{L}+X^{R}_{qt}P_{R})tZ_{\mu} + \frac{g_{w}}{4c_{w}M_{Z}} \bar{q} \sigma^{\mu \nu} (K^{L}_{qt}P_{L} + K^{R}_{qt}P_{R})tZ_{\mu \nu} + \frac{1}{\sqrt{2}} \bar{q}(\eta^{L}_{qt}P_{L} + \eta^{R}_{qt}P_{R}) tH]+ h.c.
\label{LtopFCNC}
\end{align}
in position-space for direct implementation in FeynRules \cite{feynrules}. In the UFO model the parameters can be complex numbers by letting the user define their real and imaginary parts. The UFO model can be obtained in \cite{link-ufo}. 

\section{Results}

With the UFO model we generated samples of t$\gamma$, tZ and tH events with MadGraph \cite{madgraph}. In tZ production the normalized distributions of the $p_{T}$, $\eta$, $\phi$ and energy of the top quark and Z boson as well as in the mass and energy distribution of the tZ system are not sensitive to the chiralities of the different couplings neither to their strength. However when tZ is exclusively produced through a gut, gct, Zut or Zct anomalous couplings differences in the shapes of the distributions are observed. We found analogous results for t$\gamma$ and tH production. 

\begin{figure}[h!]
\centering
\resizebox{\textwidth}{!}{
\begin{tabular}{ c c }
        \includegraphics[scale=0.5]{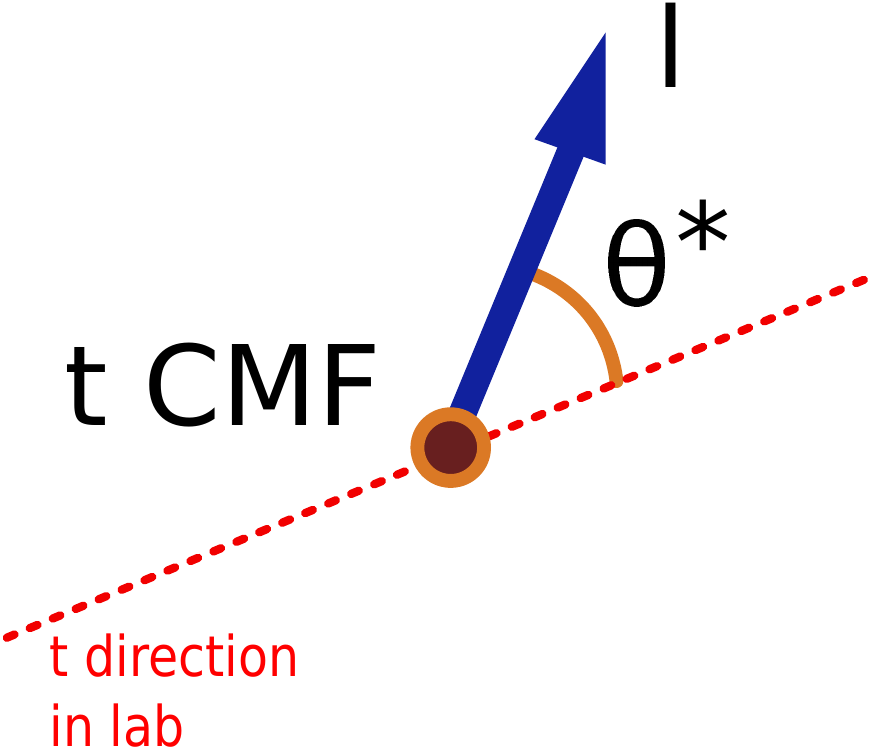} & \includegraphics[scale=0.5]{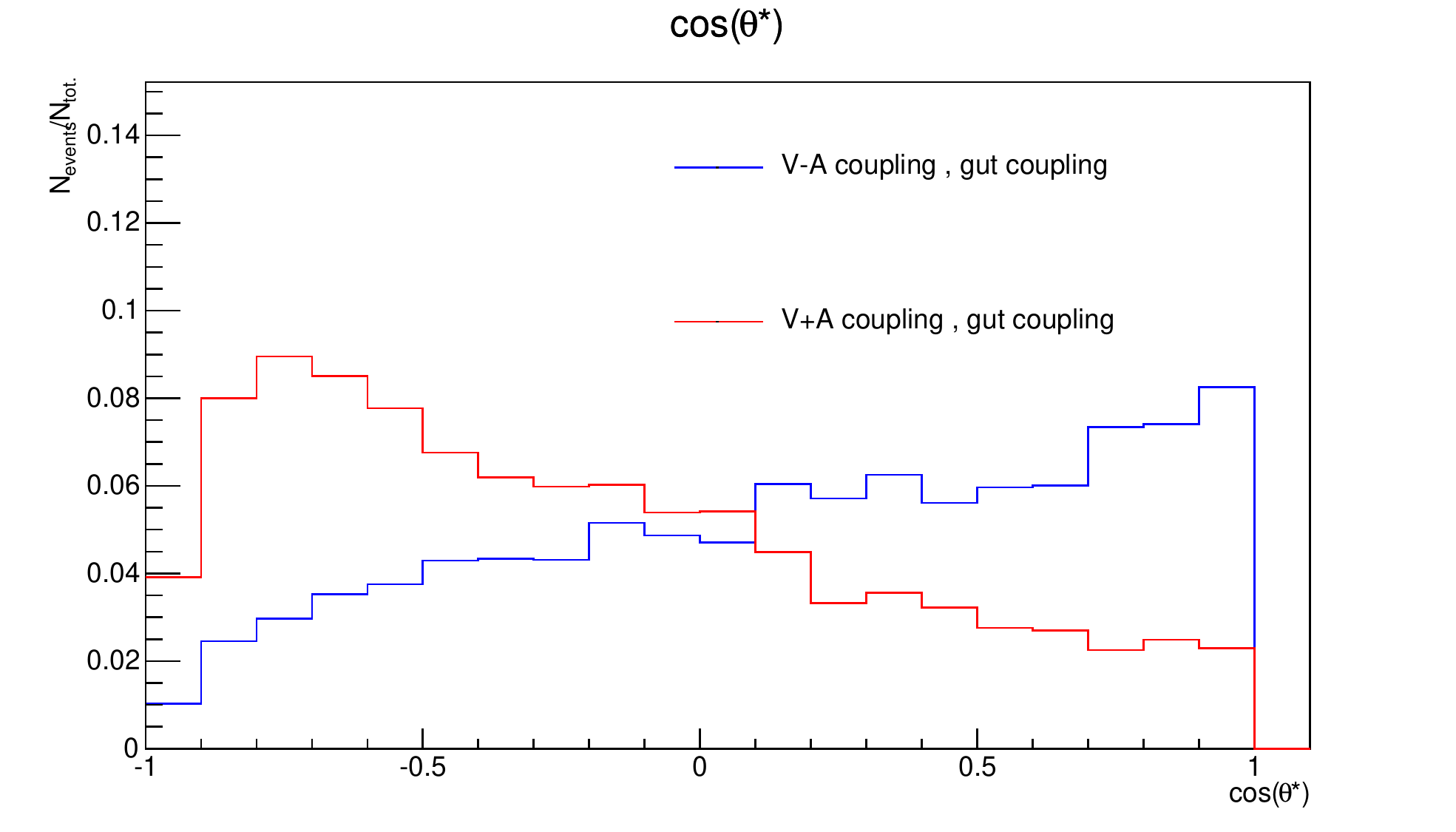} \\
        a) & b)
\end{tabular}}
\caption{a) The blue arrow is the momentum of the W lepton in the reference frame in which the top is at rest. $\theta^\ast$ is the angle between the momentum of this lepton with the direction of the top in the laboratory reference frame. b) Dependency of the $\cos (\theta^\ast)$ normalized distributions on the chirality of a gut coupling at detector-level. The results for other coupling natures and strengths are identical.}
\label{decay-products-tz}
\end{figure}

We also studied the kinematical distributions of the decay products of tZ in the trileptonic channel using a  detector level simulation with Delphes \cite{delphes} and an analysis based on the one used by CMS \cite{cms-paper}. A fair agreement was found between the obtained
events yields and those reported by CMS \cite{cms-paper}. It was observed that left-handed (LH) couplings produce b-jets with lower $p_{T}$ and leptons from the W (hereafter denoted the W leptons) with higher $p_{T}$ in comparison with right-handed (RH) couplings. Furthermore, a significant difference is observed on the angle between the W lepton momentum in the top-quark rest frame and the top-quark momentum in the laboratory frame ($\theta^*$). The fraction of events is an increasing (decreasing) function of $\cos \theta^\ast$ for LH (RH) couplings, see Figure \ref{decay-products-tz}.

A multivariate-analysis was performed in order to separate signal from background at 13 TeV. We used a Boosted Decision Tree (BDT) implemented with TMVA \cite{tmva} that was trained and tested with WZ+jets and signal samples generated with MadGraph. Half of the events of these samples were randomly selected to be used in the training phase. To train the BDT we included the discriminating variables used by CMS, except the CSV b-tagging discriminator. We also used as discriminating variables the $p_{T}$ of the b-jet and W lepton and $\cos \theta^\ast$. The trained BDT was applied to the same signal samples used to train the BDT and also to a Diboson sample. Next we performed different cuts on the BDT output followed by a counting experiment in order to extract limits on the cross-section.

In the literature it is common to find limits on the anomalous couplings $k_{gqt}/\Lambda$ and $k_{zqt}/\Lambda$ with $q=u,c$. The couplings in the Lagrangian (\ref{LtopFCNC}) can be written in terms of these anomalous couplings by the change of variables :
\begin{align}
\frac{\zeta^{R\ast}_{qt}}{2m_{t}}=\sqrt{2}\frac{k_{qct}}{\Lambda}f^{L}_{q} & \quad \text{,} & \frac{\zeta^{L\ast}_{qt}}{2m_{t}}=\sqrt{2}\frac{k_{gqt}}{\Lambda}f^{R}_{q} \quad \text{,} \\
\frac{K^{R\ast}_{qt}}{4M_{Z}}=\frac{\hat{f^{L}_{q}}}{\sqrt{2}}\frac{k_{Zqt}}{\Lambda} & \quad \text{,} & \frac{K^{L\ast}_{qt}}{4M_{Z}}=\frac{\hat{f^{R}_{q}}}{\sqrt{2}}\frac{k_{Zqt}}{\Lambda} \quad \text{,}
\end{align} 
\begin{figure}[h!]
\centering
\resizebox{\textwidth}{!}{
\begin{tabular}{ c c }
        \includegraphics[scale=0.5]{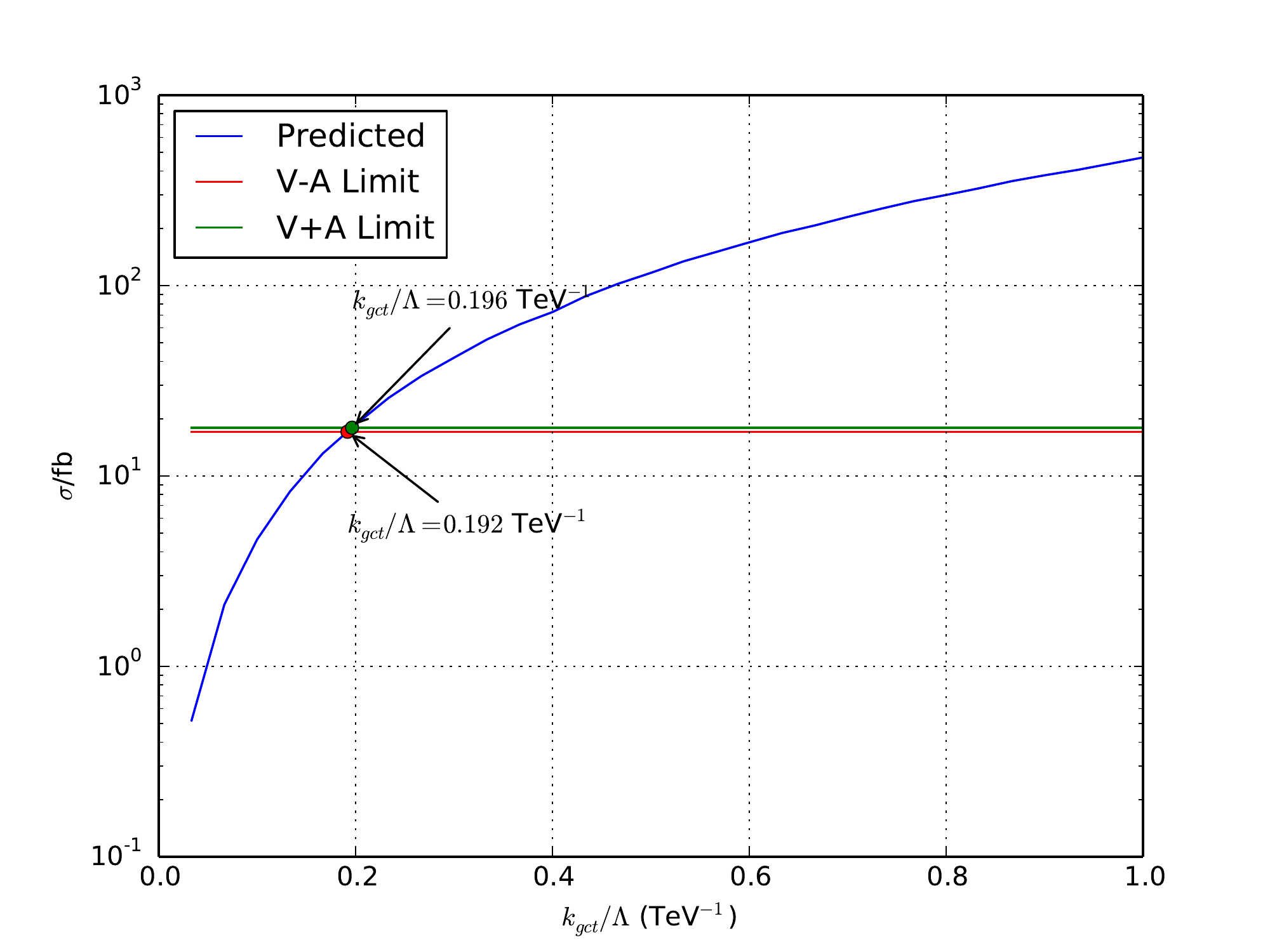} & \includegraphics[scale=0.5]{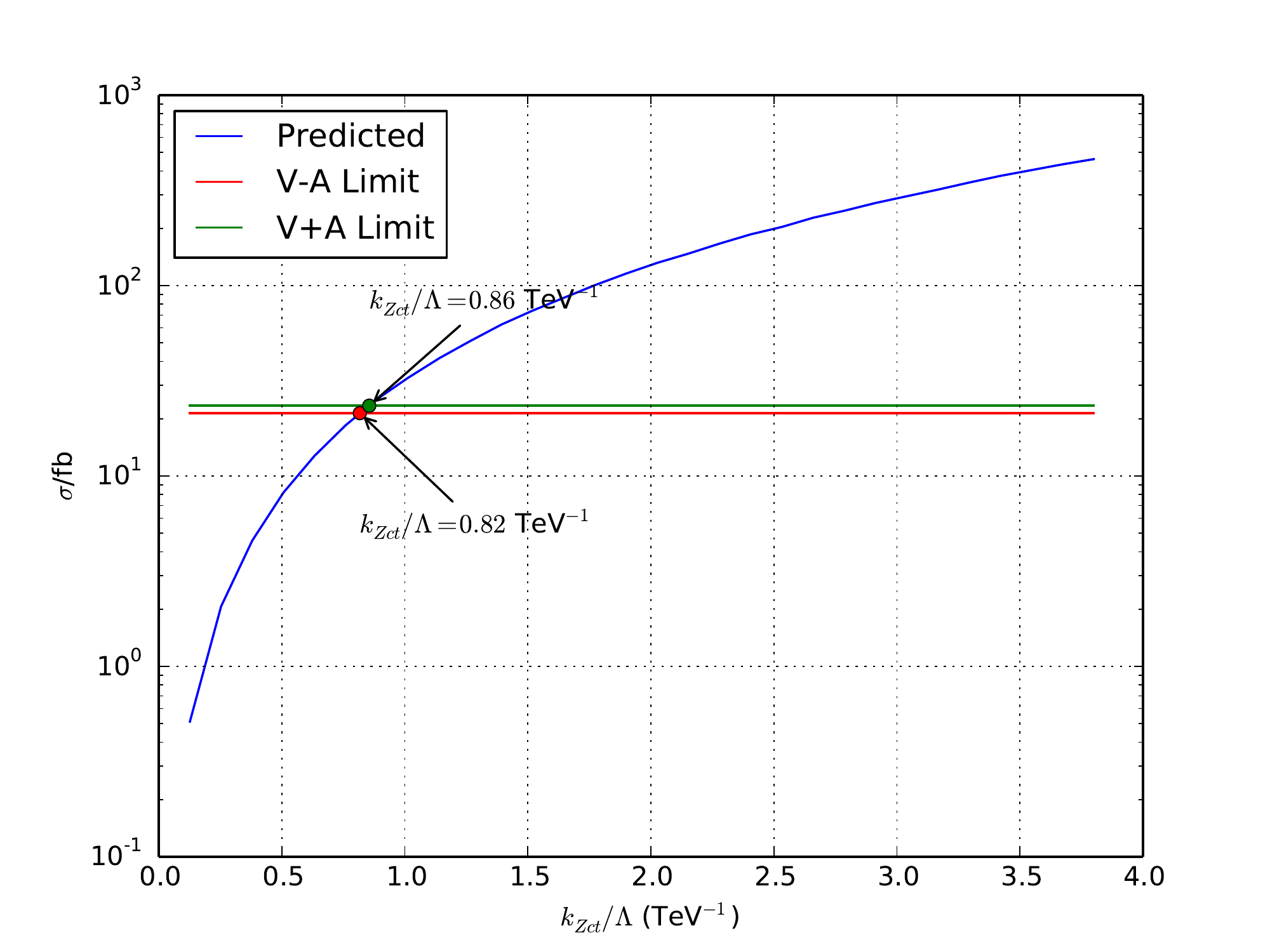} \\
    \end{tabular}}
    \label{cross-section}
    \caption{ Limits on the anomalous couplings in the case of a left-handed and right-handed coupling. The limits for gct and Zct are presented in the left and right plots respectively.}
\end{figure}
In Figure 2 you can find the upper limits at 95 \% Confidence Level for each anomalous coupling for both chiralities , asuming that only one is present at a time, and the theoretical cross-section computed with MadGraph. The limits are more constraining for a LH coupling although the difference is not significant (at the 5\% level). These results were obtained for $\sqrt{s}=13$~TeV and an integrated luminosity of 10 $\text{fb}^{-1}$.


\begin{thebibliography}{99}
\bibitem{ufo}
    C. Degrande et al. ,
    \textit{UFO - The Universal FeynRules Output},
    \emph{Comput. Phys. Commun. 183 (2012) 1201-1214},
    2012.
\bibitem{buchmuller}
    W. Buchmuller and D. Wyler,
    \emph{Effective lagrangian analysis of new interactions and flavour conservation}
    \textit{Nucl. Phys. B 268 (1986) 621}.
\bibitem{juan-top-anomalous}
    J.A. Aguilar-Saavedra,
    \textit{A minimal set of top anomalous couplings},
     \emph{Nucl.Phys.B} {\bf 812} {181-204},
     2009.
\bibitem{juan-top-higgs-anomalous}
    J.A. Aguilar-Saavedra,
    \textit{A minimal set of top-Higgs anomalous couplings},
    \emph{Nucle.Phys.B} {\bf 821} {215-227}.
\bibitem{feynrules}
    A. Alloul et al.,
    \textit{FeynRules 2.0 - A complete toolbox for tree-level phenomenology},
    \emph{Comput.Phys.Commun} {\bf 185} (2014) 2250-2300.
\bibitem{link-ufo}
    A. Amorim et al.,
    \textit{FCNC Top interactions},
    \emph{http://feynrules.irmp.ucl.ac.be/wikiGeneralFCNTop}.
\bibitem{madgraph}
    J. Alwall et al. ,
    \textit{The automated computation of tree-level and next-to-leading order differential cross sections, and their matching to parton shower simulations},
    \emph{JHEP} {\bf 07} (2014) 079,
    arXiv:1405.0301 [hep-ph].
\bibitem{delphes}
    J. de Favereau et al. ,
    \textit{DELPHES 3, A modular framework for fast simulation of a generic collider experiment},
    \emph{arXiv:1307.6346v3 [hep-ex]},
    2013. 
\bibitem{cms-paper}
    CMS Collaboration,
    \textit{Search for Flavor-Changing Neutral Currents in tZ events in proton-proton collisions at $\sqrt{s}=7$~TeV},
    CMS PAS TOP-12-021,
    2013.
\bibitem{tmva}
    A. Hocker et al. ,
    \textit{TMVA 4 - Toolkit for Multivariate Data Analysis with ROOT Users Guide},
    \emph{arXiv:physics/0703039},
    2007.
\end{thebibliography}
\end{document}